\documentclass[aps,pre,twocolumn,amssymb,showpacs]{revtex4}
\usepackage{epsfig}

\newtheorem{theorem}{Theorem}

\newcommand{\maybeeq}[1]{Eq.~(#1)}
\newcommand{\maybeeqs}[1]{Eqs.~(#1)}

\newcommand{\trace}{{\rm Trace}}
\newcommand{\id}{\mathbb I}

\begin{document}

\title{Conformal field theory correlations
in the Abelian sandpile model}

\author{M. Jeng}

\email{mjeng@siue.edu}

\affiliation{
Box 1654, Department of Physics, Southern Illinois
University Edwardsville, Edwardsville, IL, 62025}

%-----------------------------------------------------------------%

\begin{abstract}

We calculate all multipoint correlation functions
of all local bond modifications
in the two-dimensional Abelian sandpile 
model, both at the critical point, and in the model with
dissipation. The set of local bond modifications
includes, as the most physically interesting case,
all weakly allowed cluster variables.
The correlation functions show that all 
local bond modifications
have scaling dimension two, and can be written 
as linear combinations of operators in the 
central charge $-2$ logarithmic 
conformal field theory, in agreement with a form 
conjectured
earlier by Mahieu and Ruelle
in Phys. Rev. E {\bf 64}, 066130 (2001).
We find closed form expressions for
the coefficients of the operators,
and describe methods that allow their rapid calculation. 
We determine the fields associated with
adding or removing bonds, both
in the bulk, and along 
open and closed boundaries; some bond defects
have scaling dimension two, while others  
have scaling dimension four. 
We also determine the corrections
to bulk probabilities for 
local bond modifications
near open and closed boundaries.  

\end{abstract}


\maketitle

%-----------------------------------------------------------------%

\section{Introduction}

Self-organized criticality may be the underlying
cause of power laws in a wide range of natural and
man-made phenomena~\cite{BakBook,JensenBook}.
Systems exhibiting self-organized criticality
naturally approach a critical state, without any
intrinsic time or distance scales. The critical
point is reached without any fine-tuning of 
parameters. This is unlike most critical points
seen in physics---for example, the Ising model is
only critical at a single, very specific, temperature.
This lack of fine-tuning is essential if we
are to understand power laws in nature, where
no fine-tuning is possible. 

Since the concept of 
self-organized criticality was first introduced
by Bak, Tang, and Wiesenfeld in 1987, a number
of models have been developed to investigate
this phenomenon~\cite{BTW}. However, the original
model, the 
two-dimensional, isotropic,
Abelian sandpile model (ASM), is still one of
the simplest and most interesting of the models.
The ASM is simple and robust, which are
necessary features for any model of self-organized
criticality.
While natural phenomena are
quite complex, any model that seeks to explain
the ubiquity of power laws in nature must,
paradoxically, be very simple; if we are to have
a robust model for the generation of power laws,
we must neither have finely-tuned parameters,
nor finely-tuned rules.

The ASM is defined on a lattice of sites, and
is described by a toppling matrix $\bf\Delta$,
whose dimension is equal to the number of sites in
the sandpile. The sandpile evolves stochastically.
In each time step, a grain of sand is added to a random
site. Then, sites are checked for stability.
If the number of grains at a site $\vec{i}$
is greater than $\Delta_{\vec{i},\vec{i}}>0$, then the
site $\vec{i}$ is unstable, and topples, losing
$\Delta_{\vec{i},\vec{i}}$ grains, 
while every other site $\vec{j}$
gains -$\Delta_{\vec{i},\vec{j}}\geq 0$ grains.
(Generally, $\Delta_{\vec{i},\vec{j}}$ is zero
except when $\vec{j}$ 
neighbors $\vec{i}$.)
Typically, models are conservative, which means
that each
toppling in the bulk conserves the total number of 
grains ($\sum_{\vec{j}} \Delta_{\vec{i},\vec{j}} = 0$).
Only for topplings
along the boundary, where grains
can fall off the edge,
can the total number of grains change.
We continue toppling unstable sites until no
sites are unstable. Then, we begin a new time step, and
again add a grain to a random site. 

The ASM is surprisingly 
tractable~\cite{Dhar.UnitCorrelations,Dhar.First,Dhar.CFT}.
We only briefly cover some of the essential
points here---for comprehensive 
reviews, see~\cite{DharReview,Dhar.AllSame}.

After a large number of time steps, the 
ASM reaches a well-defined distribution of 
states. Of the stable height configurations, 
some are transient, and occur with probability
zero after a long amount of time. All other
states are recurrent, and 
occur with equal probability. 
Dhar showed that the total number of
recurrent states is just 
$\det({\bf \Delta})$~\cite{Dhar.First}.
This is also equal to the number of spanning
trees that can be drawn on the lattice, showing a connection between the sandpile and
spanning tree problems~\cite{Dhar.CFT}.

These statements hold for all ASMs, which define a
large class of models.
Now, we specialize to the two-dimensional,
conservative, isotropic
ASM, which is defined on a two-dimensional
square lattice, where each site has a maximum 
height of four, and where upon toppling at any site,
one grain
is sent to each of the site's
four neighbors. Furthermore,
we work in the limit where the lattice is 
infinite.
The two-dimensional, isotropic, spanning tree problem
is equivalent to the central charge -2
logarithmic conformal field theory 
($c=-2$ LCFT)~\cite{Dhar.CFT}, which has the 
simple Gaussian action 
$S=(1/\pi)\int \partial\theta\bar\partial\bar\theta$,
where $\theta$ and $\bar\theta$ are complex Grassman variables.
The $c=-2$ LCFT is described 
in~\cite{c2.Describe.1,c2.Describe.2,c2.Describe.3}.
While the two-dimensional, conservative, isotropic
ASM is just one of many possible ASMs, it is the original,
standard, model~\cite{BTW}, and it is reasonable to 
simply refer to it as ``the ASM,'' which we do for the 
remainder of this paper.

Calculations of correlation functions, using methods
to be described in the next section, have 
confirmed that there is a relationship between the
ASM and $c=-2$ LCFT. Two-point correlation functions
of unit height variables decay as $1/r^4$ in the
bulk~\cite{Dhar.UnitCorrelations}, as do 
all two-point height correlations along 
open and closed boundaries~\cite{Ivashkevich}; these
correlations can be
understood as equivalent to
correlations of LCFT operators.
Furthermore, calculations of certain 
three-point correlation functions of heights
along closed boundaries, and all multipoint
correlations of heights along
open boundaries, have allowed 
LCFT field identifications for heights
along boundaries~\cite{Mine.Short,Mine.Long}.

The ASM is not robust to all perturbations. If we relax
the constraint that the model be conservative,
and instead allow grains to be lost in any bulk
toppling (i.e. allow dissipation), correlations
decay exponentially, and we are taken off the critical
point~\cite{dissip.2,DissipBreaksSOC,dissip.1}.
The condition of conservation can be considered
a ``natural'' one, rather than one requiring ``fine-tuning.''
Deeper probes of the the conformal structure can be obtained
by looking at correlations both on and off the critical
point.

Mahieu and Ruelle calculated a number of off-critical
correlation
functions of certain height configurations,
known as weakly allowed clusters (WACs), and
used their correlation functions to propose
field identifications for the 14 simplest 
WACs~\cite{Mahieu.Ruelle}.
They found that their correlation functions could be
explained by assuming that all 14 WACs took the form

\begin{eqnarray}
\nonumber
\phi (z) & = & -
\left\{
A : \partial\theta\bar\partial\bar\theta+
    \bar\partial\theta\partial\bar\theta : +
B_1 : \partial\theta\partial\bar\theta +
      \bar\partial\theta\bar\partial\bar\theta : + \right. \\
& & \left. + i B_2 : \partial\theta\partial\bar\theta -
	\bar\partial\theta\bar\partial\bar\theta :
+{{C P(S) M^2}\over{2\pi}} 
: \theta\bar\theta :
\right\} 
\label{eq:LCFT.rep}
\end{eqnarray}

\noindent The coefficients $A$, $B_1$, $B_2$,
$C$, and $P(S)$
vary from WAC to WAC. $P(S)$ is the probability
for the cluster at the critical point,
and $M$ is the mass, a measure of how far 
the model is from 
the critical point.
The correlation functions that they used were mostly
two-point functions 
along horizontal or
diagonal axes, as well as some three-point
and four-point functions for 
the two simplest WACs.

While these calculations provide strong evidence for the
identification of the ASM with the $c=-2$ LCFT, and the
field identification in \maybeeq{\ref{eq:LCFT.rep}},
the fact that only specific correlations were considered
limits the range of the identification. It would
be surprising if 
new orientations of correlation functions, or new WACs,
were found to be
inconsistent with
\maybeeq{\ref{eq:LCFT.rep}};
but the calculations in~\cite{Mahieu.Ruelle} do not
rule this possibility out. More importantly, since
each correlation function in~\cite{Mahieu.Ruelle}
required a new and separate calculation, it is hard
to understand, mathematically, why these results occured.
While their end results showed that certain
correlations of WACs in the ASM are equal to
correlations of \maybeeq{\ref{eq:LCFT.rep}}
in the LCFT, it was not mathematically transparent 
as to why this should be.
Nor was it clear why, or if, the same coefficients would 
appear in other properties, such as off-boundary
correlations, or correlations with defects.

Here, we calculate all correlation functions of
all local bond modifications (LBMs), for 
arbitrary numbers and types of LBMs at arbitrary
positions (far from one another);
our calculations
confirm that all LBMs should receive the field 
identification in \maybeeq{\ref{eq:LCFT.rep}}.
By LBMs, we mean any set of local changes in the sandpile
toppling rules. For the ASM at the critical point, we will
assume that the LBMs are conservative (do not create or
destroy grains).
Since all WACs can be calculated by 
LBMs, our results automatically include all correlations of
WACs. While the WACs are the most important types of LBMs,
and the easiest to find probabilities of, in numerical
simulations, we generally discuss our results in terms 
of LBMs, to emphasize the generality of our results.
We give closed form expressions for $A$, $B_1$,
$B_2$, and $C,$ and describe methods that allow
rapid calculation of these
coefficients. 
While a computer
is needed for the calculation of specific
$A$, $B_1$, and $B_2$ coefficients, 
the general
calculations can be done by hand.

By showing how calculation for all LBMs can be done at
once, we make the mathematical structure
clearer. For example, we can quickly see why the
coefficients $A$, $B_1$, and $B_2$ appear in other 
properties. We illustrate this by looking at off-boundary
LBM probabilities, and correlations with bond defects
(either in the bulk, or along a boundary). Interestingly,
we find that some bond defects are represented by a
LCFT operator with a scaling dimension of four;
this is, to the best of our knowledge, the first dimension
four operator found in the ASM.

While our calculations have been done 
in both the normal ASM, and the ASM with dissipation,
we focus our discussion on
the simpler analysis at the
critical point, and only discuss the more complicated
massive correlations
in the last section, and in the appendices.

%-----------------------------------------------------------------%

\section{Weakly allowed cluster variables}

The methods used in this paper are not powerful enough to
calculate probabilities and correlations
for any height configurations. 
Even the calculation of the probability for a
site to have height two requires much more
complicated methods~\cite{Priezzhev}, and the 
correlation function of two height two variables
remains unknown. This is because the condition for a 
site to have height two involves a nonlocal condition.

\begin{figure}[tb]
\epsfig{figure=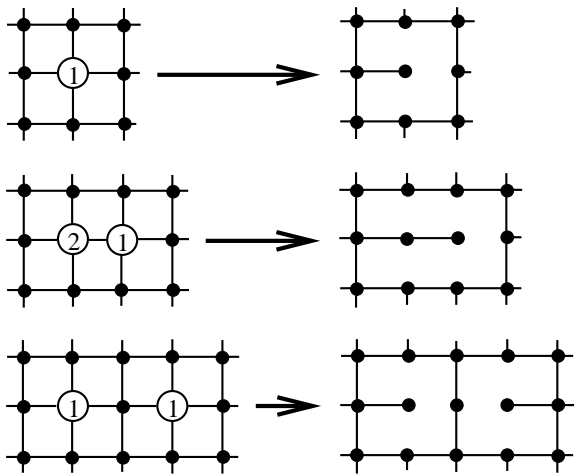,width=3.0in}
\caption{Some WACs and their corresponding modified
sandpiles}
\label{fig:WACs}
\end{figure}

As already stated, the most important LBMs
are those used to calculate properties of
WACs~\cite{Majumdar.Dhar.II}. 
WACs are related to forbidden subconfigurations (FSCs).
An FSC is a height configuration over a subset of
sites $F$, such that for every site $\vec{i}\in F$,
the number of neighbors of $\vec{i}$ in $F$ is greater
than or equal to
the height at $\vec{i}$. FSCs are important because
ASM height configurations are recurrent if and only if they
have no FSCs~\cite{Dhar.First}.
A WAC is a height configuration that contains no FSCs,
but becomes a FSC if any height in the WAC is
decreased by one. Three WACs are shown in the
left side of figure~\ref{fig:WACs}.

WACs are analytically tractable because
it turns out that the number of sandpiles
with a particular WAC is
equal to the number of recurrent states in a
sandpile with modified toppling 
rules~\cite{Dhar.UnitCorrelations}.
There are actually several different ways to modify the
toppling rules to obtain the WAC probability. 
The simplest is to,
for each connected piece of the WAC, remove
all but one of the bonds connecting it
to the rest of the lattice---the modified lattices
corresponding to the WACs are shown in
the right side of figure~\ref{fig:WACs}.
(See~\cite{Mahieu.Ruelle} for a discussion of other
ways in which the sandpile can be modified
to obtain the WAC probabilities.)
In these modified sandpiles, 
grains of sand cannot
flow along the removed bonds; to continue to 
conserve the number of grains during each toppling,
the condition for instability must be decreased
at the sites at the end of the removed bond. 
These changes result in a new toppling matrix 
$\bf \Delta'$.

As already stated, the number of recurrent states in the
ASM is $\det({\bf \Delta})$. The number of recurrent states
that have the WAC is given by $\det({\bf \Delta'})$. 
(We discuss this equivalence further in
section~\ref{sec:WAC.LBM.issues}.)
So the 
bulk probability for the WAC is given by

\begin{equation}
p = {{\det({\bf\Delta'})}\over{\det({\bf\Delta}})} =
\det (\id + {\bf BG}) \ ,
\end{equation}

\noindent where we have defined 
${\bf B}={\bf\Delta'}-{\bf\Delta}$ and
${\bf G}={\bf\Delta}^{-1}$.
$\bf G$ is the well-studied 
lattice Green function (at the critical point);
exact expressions are known for the
Green function between nearby sites, 
and asymptotic expressions for the Green function
between distant 
sites~\cite{Spitzer}.
While $\bf\Delta$, $\bf\Delta'$, and $\bf G$ all
have large dimensions (equal to the number
of sites), $\bf B$ is zero outside of
a finite collection of sites. 
When the bond between $\vec{i}$ and $\vec{j}$
is removed, $B_{\vec{i},\vec{j}}$ and
$B_{\vec{j},\vec{i}}$ are both increased by 1, while
$B_{\vec{i},\vec{i}}$ and $B_{\vec{j},\vec{j}}$
are both decreased by 1. 
For example, for the unit height probability, we have

\begin{eqnarray}
\nonumber
& & \quad\  \begin{array}{cccc}
\ \vec{i} & \ \vec{j}_1 & \ \vec{j}_2 & \ \vec{j}_3
\end{array} \\
{\bf B}  & = &
\left (
\begin{array}{cccc}
-3 & 1 & 1 & 1 \\
1 & -1 & 0 & 0 \\
1 & 0 & -1 & 0 \\
1 & 0 & 0 & -1
\end{array} \right)
\begin{array}{c}
\vec{i} \\ \vec{j}_1 \\ \vec{j}_2 \\ \vec{j}_3
\end{array}
\label{eq:B.Unit.Height}
\end{eqnarray}

\noindent  Here $\vec{i}$ is the site fixed at height
one, while $\vec{j}_1$, $\vec{j}_2$, and $\vec{j}_3$
are the three sites that $\vec{i}$ has been disconnected
from.

For any WAC,
the fact that $\bf B$ is finite-dimensional
means that the 
height probability can be found by
calculating a 
simple, finite-dimensional, matrix determinant.
All WACs thus correspond to LBMs. However, many LBMs
do not correspond to WACs. LBMs are simply any
sandpile modifications that can be modeled with a
$\bf B$ matrix that is conservative (every row and column
sums to zero) and symmetric.
Our analysis gives all
correlations of LBMs, which thus automatically
gives all correlations of WACs.

%-----------------------------------------------------------------%

\section{Correlations of Local Bond Modifications}
\label{sec:WAC.npoint}

\begin{figure}[tb]
\epsfig{figure=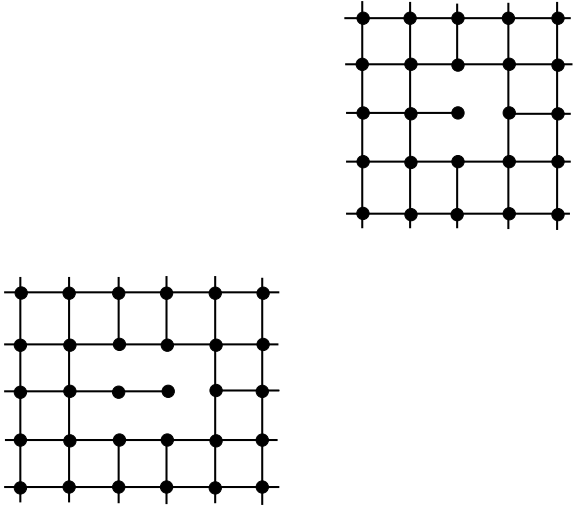,width=3.0in}
\caption{Modified sandpile for a WAC two-point correlation}
\label{fig:WAC.Correlation}
\end{figure}

For an $n$-point correlation function of LBMs, 
we can still use this method. 
The only difference is that
the removed bonds are located 
in $n$ distant clusters;
this is illustrated in figure~\ref{fig:WAC.Correlation}.
Removal of bonds in this fashion will give
${\bf B}$ and ${\bf G}$ block matrix structures.
For example, for a 3-point function, we will have

\begin{eqnarray}
{\bf B} & = &
\left(
\begin{array}{ccc}
{\bf B_1} & {\bf 0} & {\bf 0} \\
{\bf 0} & {\bf B_2} & {\bf 0} \\
{\bf 0} & {\bf 0} & {\bf B_3}
\end{array} 
\right) \\
{\bf G} & = &
\left(
\begin{array}{ccc}
{\bf G_{11}} & {\bf G_{12}} & {\bf G_{13}} \\
{\bf G_{21}} & {\bf G_{22}} & {\bf G_{23}} \\
{\bf G_{31}} & {\bf G_{32}} & {\bf G_{33}} 
\end{array}
\right)
\end{eqnarray}

\noindent ${\bf B}_u$ is
the modification to the toppling matrix
for the set of bonds removed about the $u^{\rm th}$ LBM.
${\bf G}_{uu}$ is the Green function matrix
between sites of the $u^{\rm th}$ LBM,
and its elements are $\mathcal{O}(1)$.
${\bf G}_{uv}$, $u\neq v$
is the Green function matrix between the sites
of the $u^{\rm th}$ and $v^{\rm th}$ LBMs, and its elements
are given by the bulk Green function,

\begin{equation}
\label{eq:Bulk.Green}
G_0 (x,y) = - {1\over{4\pi}} 
\ln (x^2+y^2) 
-{\gamma\over{2\pi}}
-{{\ln 8}\over{4\pi}}+\dots \ ,
\end{equation}

\noindent where $\gamma =0.57721\dots$ is the Euler-Mascheroni 
constant~\cite{Spitzer}. We work in the limit where the LBMs
are all very far from each other---we assume that any
two of the $n$ LBMs are the same order-of-magnitude,
$\mathcal{O}(r)$, apart. Since the Green function diverges
as $\ln (r)$
with increasing $r$, calculation of 
$\det (\id+{\bf BG})$ initially
looks very difficult. However, 
every row of every ${\bf B}_u$ sums to zero---this follows
from the manner in which we constructed ${\bf B}_u$,
and reflects the fact that grains of sand are still conserved
in each toppling in the bulk of the
modified sandpile. 
This implies
that parts of
${\bf G}_{uv}$ that depend only on the column
index make no contribution to $\bf BG$,
and thus no contribution to the correlation
function, $\det (\id+{\bf BG})$.
So we 
only care about differences (discrete derivatives)
of Green functions between columns
of ${\bf G}_{uv}$,
and the elements of ${\bf G}_{uv}$ are 
effectively $\mathcal{O}(1/r)$, rather than
$\mathcal{O}(\ln r)$.

For LBMs, every ${\bf B}_u$ is
symmetric, so
every column of every ${\bf B}_u$ sums to zero.
Using the matrix identity
$\det(\id +{\bf BG})=\det(\id +{\bf GB})$, this 
in turn means that the parts of 
${\bf G}_{uv}$ that depend only on the 
row index make no contribution to the probability.
This is, in effect, like taking another discrete
derivative of the Green function, so that the elements
of ${\bf G}_{uv}$ are effectively
$\mathcal{O}(1/r^2)$. 

To make this concrete, suppose that the local origin of the
$u^{\rm th}$ LBM 
is located at $(0,0)$, and the local origin of the
$v^{\rm th}$ LBM is located at
$(x_{uv},y_{uv})=(r_{uv}\cos\phi_{uv},r_{uv}\sin\phi_{uv})$.
The $u^{\rm th}$ LBM covers a set of
sites at locations
$(k_1,l_1)$, relative to $(0,0)$, 
and the $v^{\rm th}$ LBM 
consists of a series of sites at locations
$(k_2,l_2)$, relative to $(x_{uv},y_{uv})$.
($k_1$, $k_2$, $l_1$, and $l_2$ are all $\mathcal{O}(1)$.)
Then, the elements of ${\bf G}_{uv}$ all have the form
$G_0(x_{uv}+k_2-k_1,y_{uv}+l_2-l_1)$. 
The last two paragraphs show that we only need the
parts of ${\bf G}_{uv}$ that depend on {\it both}
the row {\it and} 
column indices. That is, we only need the parts of
the Green function that depend on {\it both}
$(k_1,l_1)$ {\it and} $(k_2,l_2)$, and can drop all
other terms. Expanding
\maybeeq{\ref{eq:Bulk.Green}} in powers
of $1/r_{uv}$, we find that the lowest-order term 
not dropped is

\begin{eqnarray}
\nonumber
G_0(x_{uv}+k_2-k_1,y_{uv}+l_2-l_1) & \to & 
-{1\over{2\pi r_{uv}^2}} \times
\\
& & \hspace{-2.05in} 
\left( (k_1k_2-l_1l_2) \cos (2\phi_{uv}) +
(k_1l_2+k_2l_1) \sin (2\phi_{uv}) \right) 
\nonumber \\
\label{eq:Green.r2.expansion}
\end{eqnarray}

\noindent The Green function can thus be treated
as $\mathcal{O}(1/r^2)$ for correlations of LBMs.
For more general local arrow diagrams, such as those
that appear in the calculations involving the
height two variable~\cite{Ivashkevich,Priezzhev}, 
the ${\bf B}$ matrices are not
symmetric, and we can no longer drop the parts of
${\bf G}_{uv}$ that depend only on the row index.

To get the connected $n$-point function
from $\det (\id+{\bf BG})$, we need to 
pick at least one element off the block diagonal
in every block row of $\bf G$, and in every
block column of $\bf G$. This means, at the minimum,
picking $n$ elements off the block diagonal of
$\bf G$, resulting in a leading-order 
contribution to the correlation function of 
$\mathcal{O}(1/r^{2n})$---this is the
universal part of the correlation
function. 

Mahieu and Ruelle showed that for two-point functions, the
constraint of picking only two elements off
the block diagonal allows the correlation
function to be written as~\cite{Mahieu.Ruelle}

\begin{widetext}

\begin{equation}
\label{eq:MR.Trace.2}
\det (\id+{\bf BG}) = 
-p_{u_1}p_{u_2} {\trace}
\left\{
{{\id}\over{\id+{\bf B}_{u_1}{\bf G}_{u_1u_1}}}
{\bf B}_{u_1} {\bf G}_{u_1u_2}
{{\id}\over{\id+{\bf B}_{u_2}{\bf G}_{u_2u_2}}}
{\bf B}_{u_2} {\bf G}_{u_2u_1}
\right\}
\end{equation}

\noindent Similarly, they found 
that the leading-order
contribution to the three-point probability is

\begin{eqnarray}
\nonumber
\det (\id+{\bf BG}) & = & \\
\nonumber
& &
\hspace{-0.8in}
p_{u_1}p_{u_2}p_{u_3}
{\trace}
\left\{
{{\id}\over{\id+{\bf B}_{u_1}{\bf G}_{u_1u_1}}}
{\bf B}_{u_1} {\bf G}_{u_1u_2}
{{\id}\over{\id+{\bf B}_{u_2}{\bf G}_{u_2u_2}}}
{\bf B}_{u_2} {\bf G}_{u_2u_3}
{{\id}\over{\id+{\bf B}_{u_3}{\bf G}_{u_3u_3}}}
{\bf B}_{u_3} {\bf G}_{u_3u_1}
\right\} + \\
& & 
\hspace{-0.8in}
p_{u_1}p_{u_2}p_{u_3}
{\trace}
\left\{
{{\id}\over{\id+{\bf B}_{u_1}{\bf G}_{u_1u_1}}}
{\bf B}_{u_1} {\bf G}_{u_1u_3}
{{\id}\over{\id+{\bf B}_{u_3}{\bf G}_{u_3u_3}}}
{\bf B}_{u_3} {\bf G}_{u_3u_2}
{{\id}\over{\id+{\bf B}_{u_2}{\bf G}_{u_2u_2}}}
{\bf B}_{u_2} {\bf G}_{u_2u_1}
\right\} 
\label{eq:MR.Trace.3}
\end{eqnarray}

\noindent More generally, for an $n$-point correlation,
if only $n$ terms are picked off the block diagonal, then
the connected correlation function is given by

\begin{eqnarray}
\det (\id+{\bf BG}) = {{(-)^{n+1}}\over n}
\left[ \prod_{x=1}^n p_{u_x} \right]
\sum_s
\trace \left\{
\prod_{x=1}^n 
\left[
{{\id}\over{\id+{\bf B}_{u_{s(x)}}
{\bf G}_{u_{s(x)}u_{s(x)}}}}
{\bf B}_{u_{s(x)}} {\bf G}_{u_{s(x)}u_{s(x+1)}}
\right]
\right\} \ ,
\label{eq:General.Trace.Formula}
\end{eqnarray}

\end{widetext}

\noindent where $s$ is summed over all one-to-one
mappings from  $\{1,2,\dots n\}$ to $\{1,2,\dots n\}$,
and we identify $s(n+1)$ with $s(1)$.
Mahieu and Ruelle wrote
\maybeeqs{\ref{eq:MR.Trace.2}-\ref{eq:MR.Trace.3}} in
different, but equivalent, forms. 

\begin{equation}
p_u=\det (\id+{\bf B}_u{\bf G}_{uu})
\end{equation}

\noindent is the bulk probability of the $u^{\rm th}$ LBM.
(Note that the two trace terms of 
\maybeeq{\ref{eq:MR.Trace.3}} are actually equal.
We have written the three-point function in this form to 
make clear how the form generalizes for $n$-point
functions.)

We can rewrite \maybeeq{\ref{eq:Green.r2.expansion}} as

\begin{eqnarray}
\nonumber
{\bf G}_{uv} & = &- {1\over{2\pi r_{uv}^2}} 
\left(
\left( \vec{k}_u \vec{k}_v^T - \vec{l}_u \vec{l}_v^T \right)
\cos (2\phi_{uv}) +\right. \\
& & \qquad \qquad \ 
\left. \left( \vec{k}_u \vec{l}_v^T + \vec{l}_u \vec{k}_v^T \right)
\sin (2\phi_{uv})
\right)
\label{eq:G.decompose}
\end{eqnarray}

\noindent $\vec{k}_u$ is the column vector of the horizontal
positions of the sites of the $u^{\rm th}$ LBM,
relative to the $u^{\rm th}$ local origin (i.e., 
the elements of $\vec{k}_u$
are $\mathcal{O}(1)$). $\vec{l}_u$ is the corresponding
vector of vertical positions.
$\vec{k}_u$ and $\vec{l}_u$
are both length $N_u$, where $N_u$
is the number of sites needed to represent
the $u^{\rm th}$ LBM with the methods of
the previous section
(e.g. $N_u=4$ for the unit height variable).

We insert \maybeeq{\ref{eq:G.decompose}} into
\maybeeq{\ref{eq:General.Trace.Formula}}.
For each of the $n$ ${\bf G}_{uv}$'s, we can pick any of the
four matrices of
\maybeeq{\ref{eq:G.decompose}},
resulting in $4^n$ terms. In each of these $4^n$
terms,
each ${\bf G}_{uv}$ has been replaced with
the product of a 
column vector and a row vector.
Using the cyclicity of the trace to move one
row vector at the end of the trace to the start of
the trace,
we see that each matrix 
$(\id+{\bf B}_u{\bf G}_{uu})^{-1} {\bf B}_u$
is bracketed by a row vector to its left, and a column
vector to its right, producing a $1\times 1$
matrix. So each of the $4^n$ terms is
the product of $n$ numbers. We can represent the
decisions as to which terms of 
\maybeeq{\ref{eq:G.decompose}} to pick by representing
${\bf G}_{uv}$ with a $2\times 2$ matrix, ${\bf N}_{uv}$. 
The possible ways to bracket
$(\id+{\bf B}_u{\bf G}_{uu})^{-1} {\bf B}_u$
can be represented with a $2\times 2$ 
matrix, ${\bf M}_u$.
We have

\begin{eqnarray}
\nonumber
& & \ \ \ \ \ 
\vec{k}_u^T \quad\  \vec{l}_u^T
\\
\label{eq:Crit.M.matrix}
{\bf M}_u & \equiv &
\left(
\begin{array}{cc}
c_{u,kk} & c_{u,kl} \\
c_{u,kl} & c_{u,ll}
\end{array} 
\right) 
\begin{array}{c}
\vec{k}_u \\
\vec{l}_u
\end{array}
\\
\nonumber
& & \qquad\qquad\qquad\ 
\vec{k}_v^T \qquad\qquad \vec{l}_v^T
\\
\label{eq:Crit.N.matrix}
{\bf N}_{uv} & \equiv &
- {1\over{2\pi r_{uv}^2}}
\left(
\begin{array}{cc}
\cos (2\phi_{uv}) & \sin(2\phi_{uv}) \\
\sin (2\phi_{uv}) & -\cos(2\phi_{uv})
\end{array} 
\right) 
\begin{array}{c}
\vec{k}_u \\
\vec{l}_u
\end{array}
\end{eqnarray}

\noindent We have defined

\begin{eqnarray}
\label{eq:cu.kk}
c_{u,kk} & \equiv & 
-p_u
\vec{k}_u^T
{\id\over{\id+{\bf B}_u{\bf G}_{uu}}} {\bf B}_u
\vec{k_u} \\
\nonumber
c_{u,kl} & \equiv & 
-p_u
\vec{k}_u^T
{\id\over{\id+{\bf B}_u{\bf G}_{uu}}} {\bf B}_u
\vec{l_u} \\
& = & 
-p_u
\vec{l}_u^T
{\id\over{\id+{\bf B}_u{\bf G}_{uu}}} {\bf B}_u
\vec{k_u} \\
\label{eq:cu.ll}
c_{u,ll} & \equiv & 
-p_u
\vec{l}_u^T
{\id\over{\id+{\bf B}_u{\bf G}_{uu}}} {\bf B}_u
\vec{l_u}
\end{eqnarray}

Then, the correlation function of $n$ LBMs is given by

\begin{eqnarray}
\nonumber
-\trace\left( {\bf M}_{u_1} {\bf N}_{u_1u_2} 
{\bf M}_{u_2} {\bf N}_{u_2u_3} \dots
{\bf M}_{u_n} {\bf N}_{u_nu_1}\right) & & \\
& & \hspace{-2.3in}
- \left( ((n-1)!-1)\ {\rm other\ trace\ terms} \right)\ ,
\label{eq:npt.Trace.Form}
\end{eqnarray}

\noindent where the other trace terms are derived
by permutations of $\{u_1,u_2,\dots,u_n\}$, as in 
\maybeeq{\ref{eq:General.Trace.Formula}}.

We can compare this to correlation functions
of fields in the $c=-2$ LCFT. Mahieu and Ruelle 
proposed that the
WACs are represented, at the critical point, by

\begin{eqnarray}
\nonumber
\phi_u (z_u) & = & -
\left\{
A_u : \partial\theta\bar\partial\bar\theta+
    \bar\partial\theta\partial\bar\theta : +
B_{1u} : \partial\theta\partial\bar\theta+
      \bar\partial\theta\bar\partial\bar\theta :  \right. \\
& & \left. \qquad + i B_{2u} : \partial\theta\partial\bar\theta -
	\bar\partial\theta\bar\partial\bar\theta :
\right\}
\label{eq:LCFT.rep.critical}
\end{eqnarray}

\noindent (The ``C'' term in
\maybeeq{\ref{eq:LCFT.rep}} only appears
off the critical point.)

We can compute connected
$n$-point correlations of these fields
in the $c=-2$ LCFT. We use the 
formulation of the $c=-2$ LCFT where the action is

\begin{equation}
S={1\over\pi} \int d^2x :\partial\theta\bar\partial\bar\theta: 
\ \ ,
\label{eq:LCFT.action}
\end{equation}

\noindent where we don't integrate over zero modes in
expectation values.
Since the theory is Gaussian, to
calculate correlation functions
we simply need to take Wick contractions.
The relevant nonzero ones are

\begin{eqnarray}
\label{eq:crit.contraction.1}
\left< \partial\theta(z_u) \partial\bar\theta(z_v) \right> & = &
-{1\over {2 (z_u-z_v)^2}} = 
-{{e^{-2i\phi_{uv}}}\over{2r_{uv}^2}} \\
\label{eq:crit.contraction.2}
\left< \bar\partial\theta(z_u) \bar\partial\bar\theta(z_v) \right> & = &
-{1\over {2 (\bar{z}_u-\bar{z}_v)^2}} = 
-{{e^{+2i\phi_{uv}}}\over{2r_{uv}^2}} 
\end{eqnarray}

Each term of \maybeeq{\ref{eq:LCFT.rep.critical}}
has one $\theta$, and one $\bar\theta$. The only 
difference between terms is whether the derivative on
the $\theta$ is holomorphic or antiholomorhic, and
whether the derivative on the $\bar\theta$ is holomorphic or
antiholomorphic. We can 
use a $2\times 2$ matrix
to represent the choice of which
terms of $\phi_u (z_u)$ are picked:

\begin{eqnarray}
\nonumber
& & \qquad\qquad\ \begin{array}{cc}
\partial\bar\theta & \qquad\quad\ \ \ \bar\partial\bar\theta \\
\end{array} \\
{\bf F}_u  & = &
\begin{array}{c}
\partial\theta \\ \bar\partial\theta
\end{array}
\left (
\begin{array}{cc}
B_{1u}+iB_{2u} & A_u \\
A_u & B_{1u}-iB_{2u}
\end{array} \right)
\label{eq:F.critical}
\end{eqnarray}

\noindent The contractions of
\maybeeqs{\ref{eq:crit.contraction.1}-\ref{eq:crit.contraction.2}}
can then be represented with the matrix

\begin{eqnarray}
\nonumber
& & \qquad\qquad\  \begin{array}{cc}
\quad\partial\theta & \qquad\qquad\qquad \bar\partial\theta \\
\end{array} \\
{\bf H}_{uv}  & = &
\begin{array}{c}
\partial\bar\theta \\ \bar\partial\bar\theta
\end{array}
\left (
\begin{array}{cc}
-e^{-2i\theta_{uv}}/(2r_{uv}^2) & 0 \\
0 & -e^{+2i\theta_{uv}}/(2r_{uv}^2) 
\end{array} \right)
\nonumber \\
\label{eq:H.critical}
\end{eqnarray}

\noindent The contribution to the correlation function 
where the $\bar\theta$ of the 1st LBM contracts
with the $\theta$ of the 2nd LBM, the $\bar\theta$
from the 2nd LBM contracts with the $\theta$ of the
3rd LBM, and so on,  is

\begin{equation}
-\trace (
{\bf F}_{u_1} {\bf H}_{u_1u_2} {\bf F}_{u_2} {\bf H}_{u_2u_3} \dots
{\bf F}_{u_n} {\bf H}_{u_nu_1}
)
\label{eq:LCFT.critical.trace}
\end{equation}

\noindent Other contractions give other permutations, just
as in \maybeeq{\ref{eq:npt.Trace.Form}}. Finally, 
${\bf M}_u{\bf N}_{uv}$
differs from ${\bf F}_u{\bf G}_{uv}$ only by a matrix
rotation, which will not affect the trace, if we make the 
following identifications:

\begin{eqnarray}
\label{eq:A}
A_u & = & {1\over{2\pi}} (c_{u,kk}+c_{u,ll}) \\
B_{1u} & = & {1\over{2\pi}} (c_{u,kk}-c_{u,ll}) \\
\label{eq:B2}
B_{2u} & = & {1\over\pi} c_{u,kl}
\end{eqnarray}

\noindent 
So the traces in \maybeeq{\ref{eq:npt.Trace.Form}}
and \maybeeq{\ref{eq:LCFT.critical.trace}} 
are equal, and
all LBMs are indeed represented 
by the field in \maybeeq{\ref{eq:LCFT.rep.critical}}. 
These formulas for the coefficients have the
appropriate transformation properties
under $90^{\rm O}$ rotations, and x and y 
reflections.
(Technically, the overall sign of 
\maybeeq{\ref{eq:A}} is still
undetermined at
this point, since all correlation functions have even
numbers of $A$'s. To determine the signs of the $A$'s
we need to look at at least one massive correlation
function. We can do this by consulting the
massive three-point function of the unit height variable
in~\cite{Mahieu.Ruelle}, or more broadly, by
looking at the general massive correlations in
section~\ref{sec:Dissipation.General}.)

%-----------------------------------------------------------------%

\section{Computation of \protect\boldmath$A$, $B_1$ and $B_2$ terms}
\label{sec:Computation.S10}

$A$, $B_1$, and $B_2$, can be calculated 
on a computer with
\maybeeqs{\ref{eq:A}-\ref{eq:B2}} and
\maybeeqs{\ref{eq:cu.kk}-\ref{eq:cu.ll}}.
Evaluating \maybeeqs{\ref{eq:cu.kk}-\ref{eq:cu.ll}},
as written,
requires taking a matrix inverse, which can
be computationally
time-consuming for larger LBMs. The calculation can
be made substantially faster with
the following matrix identity, which
we state without proof:

\begin{eqnarray}
\nonumber
\det\left(\id+{\bf B}\left({\bf G}+K
\vec{f}\vec{g}^T\right)\right) & = &
\det\left(\id+{\bf BG}\right)+\\
& & \hspace{-1.2in}
+ K \det\left( \id+{\bf BG}\right)
\left(
\vec{g}^T {\id\over{\id+{\bf BG}}} {\bf B} \vec{f}
\right)
\label{eq:if.G.factorizes}
\end{eqnarray}

\noindent This identity holds for
any vectors $\vec{f}$ and $\vec{g}$,
and c-number $K$.
It allows us to compute the $c$'s in
\maybeeqs{\ref{eq:cu.kk}-\ref{eq:cu.ll}}
as matrix determinants, which 
is faster than computing matrix inverses.
Furthermore, we note that, in general,
certain combinations of rows (and columns)
of ${\bf B}_u$ will sum to zero,
which means that we can perform a matrix rotation to reduce
the size of the matrix determinant. 
With these methods, computation of 
$A$, $B_1$, and $B_2$ for the ten simplest WACs takes
roughly one hour, using Mathematica on a computer with
a 1.2 GHz processor. The results agree with those found
in~\cite{Mahieu.Ruelle}. Comparison with two of
larger WACs, which they label $S_{10}$ and $S_{11}$,
requires a more detailed discussion of the mapping
between WACs and LBMs, which is done in the next section.

%-----------------------------------------------------------------%

\section{The mapping between weakly allowed cluster and
local bond modifications}
\label{sec:WAC.LBM.issues}

\begin{figure}[tb]
\epsfig{figure=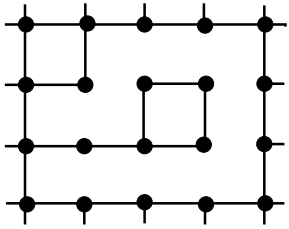,width=1.5in}
\caption{Modified sandpile for the height
configurations in figures~\ref{fig:S10.MR} and
\ref{fig:S10.Actual.WACs}}
\label{fig:S10.Lattice.Modification}
\end{figure}

\begin{figure}[tb]
\epsfig{figure=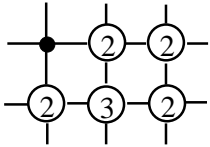,width=1.1in}
\caption{A non-WAC with the same shape as the configuration
in figure~\ref{fig:S10.Lattice.Modification}}
\label{fig:S10.MR}
\end{figure}

\begin{figure}[tb]
\epsfig{figure=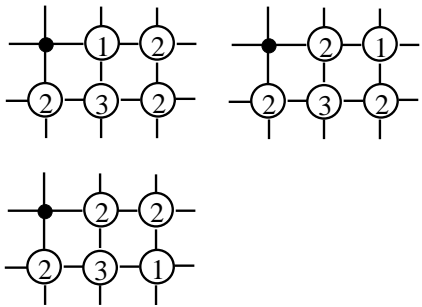,width=2.2in}
\caption{Three WACs with the same shape as 
the configuration in figure~\ref{fig:S10.Lattice.Modification}}
\label{fig:S10.Actual.WACs}
\end{figure}

We illustrate the mapping between larger weakly allowed
clusters and local bond modifications with the 
sandpile modification shown in
figure~\ref{fig:S10.Lattice.Modification}.
In this modified sandpile, a five-site cluster is separated
from the rest of the sandpile, except by a single bond.
The number of states in the modified sandpile
of figure~\ref{fig:S10.Lattice.Modification}
is equal to the number of states of the unmodified
ASM where decreasing the five-site cluster's
leftmost site (which we call $\vec{i}$)
from 2 to 1 makes the five-site
cluster a FSC, {\it and} does not produce
any larger FSCs~\cite{Priezzhev}.
The condition that the FSC produced be maximal
is necessary for this equivalence,
although this condition was not explicitly
stated in~\cite{Priezzhev}.
If changing the height of $\vec{i}$ from 2 to 1
makes the five-site cluster a FSC, the original
height configuration 
(before this change)
must have been one of the four configurations
shown in figures~\ref{fig:S10.MR}
and~\ref{fig:S10.Actual.WACs}. Of these four
configurations, the one in
figure~\ref{fig:S10.MR} is not a WAC, while the 
three in
figure~\ref{fig:S10.Actual.WACs} are.

Absent other conditions, the configuration in 
figure~\ref{fig:S10.MR} does not have the same 
probability as the configurations in
figure~\ref{fig:S10.Actual.WACs}. The configuration
in figure~\ref{fig:S10.MR} is more likely, as
an allowed configuration always stays allowed when
a site height is increased. However, for some 
configurations of heights
outside the five-site cluster, decreasing one of the three
height two sites other than the one at $\vec{i}$, to height
one,
will create an FSC outside the five-site cluster, so that
figure~\ref{fig:S10.MR} is allowed, but 
figure~\ref{fig:S10.Actual.WACs} is not.

However, if we impose the condition that taking the height
of $\vec{i}$ from two to one should produce no FSC larger than the
five-site cluster, then all four configurations
in figure~\ref{fig:S10.MR} and~\ref{fig:S10.Actual.WACs}
are equally probable. With this condition, decreasing one
of the height two sites in figure~\ref{fig:S10.MR} cannot
possibly create an FSC outside the five-site cluster, since
then the union of this FSC with the five sites would be
a larger FSC upon decreasing the height of $\vec{i}$.

For a WAC, it can be shown that if any height
is decreased, not only does the WAC become a FSC, but
it is not contained in any larger FSC. Therefore,
for the three height configurations in 
figure~\ref{fig:S10.Actual.WACs}, the condition that
the five-site FSC generated is maximal is automatic.
(However, for the configuration in 
figure~\ref{fig:S10.MR} it is not.)
Therefore, the probability associated with
the modified sandpile in
figure~\ref{fig:S10.Lattice.Modification} is
four times the probability of any of the three
WACs in figure~\ref{fig:S10.Actual.WACs}. It is also
four times the probability of the height configuration
in figure~\ref{fig:S10.MR}, if we impose on this
configuration the condition that
decreasing the height of $\vec{i}$ should produce
no FSC larger than the five-site cluster (although 
clearly this is not as physically interesting).

In~\cite{Mahieu.Ruelle}, the configurations that they
labelled $S_{10}$ and $S_{11}$ were not WACs. Once they are
modified to be WACs with the same shape,
our values for $A$, $B_1$, and $B_2$,
obtained with the methods of the previous section,
agree with theirs.

Although we chose a specific sandpile modification,
the discussion is easily generalized. Generally, 
consider sandpile
modifications similar to those in
figure~\ref{fig:S10.Lattice.Modification},
which separate a cluster of sites from the rest of the
sandpile, except for one linking site.
There will be $N$ possible height
configurations in the cluster that become FSCs if the
height of the linking site is reduced.
Some of these will be WACs and some will not. The
probability associated with the lattice modification will
be $N$ times the probability for any individual WAC.
(It will also be $N$ times the probability for any 
of the other height
configurations, given the condition that
decreasing the height of the link site should produce
no FSC larger than the cluster in question.)

These issues did not need to be discussed 
for the simpler WACs, such as those in
figure~\ref{fig:WACs}
(or the other WACs of~\cite{Mahieu.Ruelle}).
For each of the sandpile modifications in the
right side of figure~\ref{fig:WACs}, only one
corresponding WAC height configuration is 
possible ($N=1$).

%-----------------------------------------------------------------%

\section{Off-boundary expectation values}

These methods allow
us to quickly determine the effects of a number of
defects or boundary conditions on all
LBMs, and see that each time, we obtain the same
coefficients. In many cases,
a defect or boundary changes the Green
function matrix in a manner such that the change factorizes,
taking the form
$K\vec{f}\vec{g}^T$, for some
vectors $\vec{f}$ and $\vec{g}$.
\maybeeq{\ref{eq:if.G.factorizes}}
then shows that the effects of the
change can be written as a
linear combination of the $c$'s, and thus as a
linear combination of $A$, $B_1$, and $B_2$.

For example, consider the probability for an LBM 
located a distance $y$ from a boundary (open or closed).
Let the boundary be at $y=0$, and $x$ be the coordinate
along the boundary. Then the Green function is modified 
to~\cite{Bdy.Falloff}

\begin{eqnarray}
\nonumber
G_{\rm open}(x_1,y_1;x_2,y_2) & = &
G_0(x_1-x_2,y_1-y_2) \\
& & \hspace{-0.2in}
-G_0(x_1-x_2,y_1+y_2+2) 
\label{eq:Green.open} \\
\nonumber
G_{\rm closed}(x_1,y_1;x_2,y_2) & = &
G_0(x_1-x_2,y_1-y_2)\\
& & \hspace{-0.2in}
+G_0(x_1-x_2,y_1+y_2+1)
\label{eq:Green.closed}
\end{eqnarray}

\noindent Placing the local origin of the LBM at (0,y), the
Green function between points $(k_1,l_1)$ and $(k_2,l_2)$,
relative to this local origin, where $k_1$, $l_1$, $k_2$ and
$l_2$ are all $\mathcal{O}(1)$, is 

\begin{eqnarray}
\nonumber
G(k_1,l_1;k_2,l_2) & =  & G_0(k_1-k_2,l_1-l_2) 
\pm {{k_1k_2+l_1l_2}\over{8\pi y^2}} \\
& & + \mathcal{O} \left( {1\over{y^3}} \right)
\end{eqnarray}

\noindent In the $\pm$, the top sign
is for closed boundaries, and the bottom sign
is for open boundaries. 
As with \maybeeq{\ref{eq:Green.r2.expansion}},
we have only kept terms that depend on
both $(k_1,l_1)$ and $(k_2,l_2)$
(otherwise, there would be terms of
$\mathcal{O}(\ln y)$ and $\mathcal{O}(1/y)$).
Then, using
\maybeeq{\ref{eq:if.G.factorizes}}, keeping only
terms of $\mathcal{O}(1/y^2)$, and using the definitions in
\maybeeqs{\ref{eq:A}-\ref{eq:B2}} and
\maybeeqs{\ref{eq:cu.kk}-\ref{eq:cu.ll}}, we
immediately see that the probability for the LBM is

\begin{eqnarray}
p_u \mp {A_u \over {4y^2}} + \mathcal{O} \left({1\over y^3}\right)
\end{eqnarray}

\noindent This agrees with results for the unit height
variable found in~\cite{Bdy.Falloff}.

%-----------------------------------------------------------------%

\section{Bond defects in the bulk}
\label{sec:bond.bulk}

We have also used the methods described here to investigate
bond defects in the ASM. We change the toppling matrix
from the defect-free matrix ${\bf\Delta}_0$ to

\begin{eqnarray}
\Delta (\vec{i},\vec{j}) & = & \Delta_0 (\vec{i},\vec{j}) +
\delta\Delta (\vec{i},\vec{j}) \\
\delta\Delta (\vec{i},\vec{j}) & = &
\left\{
\begin{array}{cc}
-k_{\rm bond} & {\rm if}\ 
(\vec{i},\vec{j})=(\vec{s}_0,\vec{s}_0)\ {\rm or}\\
& \quad 
(\vec{i},\vec{j})=(\vec{s}_1,\vec{s}_1)\\
+k_{\rm bond} & {\rm if}\ 
(\vec{i},\vec{j})=(\vec{s}_0,\vec{s}_1)\ {\rm or}\\
& \quad
(\vec{i},\vec{j})=(\vec{s}_1,\vec{s}_0)
\end{array} 
\right. 
\end{eqnarray}

\noindent If $\vec{s}_0$ and $\vec{s}_1$ are adjacent
sites, and $k_{\rm bond}=1$, then this corresponds to
removing the bond between $\vec{s}_0$ and $\vec{s}_1$.
If $k_{\rm bond}=-1$, this corresponds to adding
a bond between $\vec{s}_0$ and $\vec{s}_1$.

The Green function is the inverse of the toppling matrix,
and the effects of this perturbation can 
be calculated by summing a geometric series.
The result is

\begin{eqnarray}
\nonumber
G(\vec{i},\vec{j}) & = & G_0 (\vec{i},\vec{j}) +
\tilde{k}_{\rm bond}
(G_0(\vec{i},\vec{s}_0)-G_0(\vec{i},\vec{s}_1)) \times \\
& & \qquad\qquad\qquad\ 
(G_0(\vec{s}_0,\vec{j})-G_0(\vec{s}_1,\vec{j})) \ ,
\label{eq:Green.bond}
\end{eqnarray}

\noindent where

\begin{eqnarray}
\nonumber
{1\over{\tilde{k}_{\rm bond}}}-{1\over{k_{\rm bond}}} & = &
\\
\nonumber
& & \hspace{-1.0in}
G_0(\vec{s}_0,\vec{s}_0)+G_0(\vec{s}_1,\vec{s}_1)-
 G_0(\vec{s}_0,\vec{s}_1)-G_0(\vec{s}_1,\vec{s}_0)
\\
\end{eqnarray}

\noindent The correction to the Green function 
factorizes, just as 
in \maybeeq{\ref{eq:if.G.factorizes}},
and the corrections to the LBM probabilities
again depend on the same coefficients.
If the two ends of the bond defect are at $(0,0)$ and
$(q_x,q_y)$, where $q_x$ and $q_y$ are both
$\mathcal{O}(1)$, then from
\maybeeqs{\ref{eq:cu.kk}-\ref{eq:cu.ll}},
\maybeeqs{\ref{eq:A}-\ref{eq:B2}}, and
\maybeeq{\ref{eq:if.G.factorizes}},
the probability for a LBM of type $u$
at $(x,y)=(r\cos\theta,r\sin\theta )$, for $r\gg 1$, is

\begin{eqnarray}
\nonumber
-{{\tilde{k}_{\rm bond}}\over{4\pi r^4}} & &
\!\!\!\!\!\!
\left\{
(q_x^2+q_y^2)A_u + \right. \\
\nonumber
& &
(q_x^2-q_y^2)(B_{1u}\cos (4\theta)+B_{2u}\sin (4\theta))+  \\
& &
\left. (2 q_x q_y) (B_{1u}\sin (4\theta)-B_{2u}\cos(4\theta))
\right\}
\end{eqnarray}

\noindent This is consistent with identifying the bond
defect with

\begin{equation}
-{{\tilde{k}_{\rm bond}} \over {2\pi}}
: ( q_x \partial_x \theta + q_y \partial_y \theta )
  ( q_x \partial_x \bar\theta + q_y \partial_y \bar\theta ) :
\end{equation}

%-----------------------------------------------------------------%

\section{Boundary operators and bond defects}

It was shown in~\cite{Mine.Long} that any local arrow
diagram along an open boundary is represented 
in the LCFT by the operator

\begin{eqnarray}
\nonumber
& & -{2\over\pi} \det(\id+{\bf B}_u{\bf G}_{uu})
\left( (\vec{y}+\vec{1})^T 
{\id\over{\id+{\bf B}_u{\bf G}_{uu}}} {\bf B}_u
(\vec{y}+\vec{1}) \right) \\
& &
\qquad\qquad \times
\partial\theta\partial\bar\theta\ ,
\label{eq:open.general.field}
\end{eqnarray}

\noindent where $\vec{y}$ is the vector of distances
perpendicular to the boundary. The arguments there worked
for {\it any} local arrow diagram
(not just those corresponding to LBMs), and 
similarly to our arguments in
section~\ref{sec:WAC.npoint},
used the fact
that the Green function along open boundaries falls off 
as $1/x^2$. Along closed boundaries (and in the bulk),
the Green function grows as $\ln x$, so the situation
is more complicated, and not all operators are
proportional to $\partial\theta\partial\bar\theta$.
(See, for example, the operators for the height
two and three variables along closed boundaries, given 
in~\cite{Mine.Short,Mine.Long}.)
However, we can now derive an expression similar
to \maybeeq{\ref{eq:open.general.field}}
for LBMs along closed boundaries.

The Green function for two sites on a closed boundary, and
$x\gg 1$ apart, was found in~\cite{Bdy.Falloff},
using \maybeeq{\ref{eq:Green.closed}}, to be

\begin{eqnarray}
\nonumber
G_{\rm closed}(0,0;x,0) & = &
-{1\over\pi} \ln (x)
-\left( {\gamma\over\pi} +{{\ln 2}\over{2\pi}}\right)
+{1\over{6\pi x^2}}  \\
& & 
+\mathcal{O}\left({1\over{x^4}}\right)
\label{eq:Green.closed.expansion.1}
\end{eqnarray}

\noindent Using the recursion relation
${\bf G\Delta}=\id$
on this equation, we can 
extend \maybeeq{\ref{eq:Green.closed.expansion.1}}
for points $\mathcal{O}(1)$ from the 
boundary:

\begin{eqnarray}
\nonumber
G_{\rm closed}(x_1,y_1;x+x_2,y_2) & =  & G_{\rm
closed}(0,0;x,0) \\
& & \nonumber \hspace{-1.5in}
-{1\over\pi} \ln \left( 1+ {{x_2-x_1}\over x}\right)
-{{y_1(y_1+1)+y_2(y_2+1)}\over{2\pi (x+x_2-x_1)^2}} \\
& & \hspace{-1.5in}
+{{x_2-x_1}\over{3\pi x^3}}+
\mathcal{O} \left( {1\over{x^4}} \right)
\label{eq:Green.closed.expansion.2}
\end{eqnarray}

\noindent ($x_1$, $x_2$, $y_1$, and $y_2$
are all $\mathcal{O}(1)$.)
The Green function diverges as
$\ln x$, but
if we are calculating correlations of 
LBMs along closed boundaries, we can use the arguments
of section~\ref{sec:WAC.npoint} to see that we only
care about the parts of the Green function matrix
that depend on both the row and the column indices.
The part of 
\maybeeq{\ref{eq:Green.closed.expansion.2}} 
that depends on both
$(x_1,y_1)$ and $(x_2,y_2)$ is

\begin{eqnarray}
\nonumber
G_{\rm closed}(x_1,y_1;x+x_2,y_2) & \to &
- {{x_1x_2}\over{\pi x^2}} \\
& & \nonumber \hspace{-1.5in}
+ {{x_1^2x_2-x_1x_2^2+x_1y_2(y_2+1)-x_2y_1(y_1+1)}
\over{\pi x^3}} \\
& & \hspace{-1.5in}
+\mathcal{O}\left({1\over {x^4}}\right)
\end{eqnarray}

\noindent Using logic identical to that in the bulk case,
we can use the $\mathcal{O}(1/x^2)$ part, to derive
field identifications
for LBMs along closed
boundaries:

\begin{equation}
{2\over\pi} \det(\id+{\bf B}_u{\bf G}_{uu})
\left( \vec{x}^T 
{\id\over{\id+{\bf B}_u{\bf G}_{uu}}} {\bf B}_u
\vec{x} \right)
\partial\theta\partial\bar\theta 
\end{equation}

\noindent $\vec{x}$ 
is the vector of position coordinates parallel to the
boundary. 

We now introduce a bond defect 
of strength $k_{\rm bond}$,
along an open or closed boundary,
between sites
$(q_{x1},q_{y1})$ and $(q_{x2},q_{y2})$, that
are $\mathcal{O}(1)$ apart. 
These bond defects can be analyzed 
as in the previous section.
Along an
open boundary, the bond defect is represented by

\begin{equation}
{2\over\pi} \tilde{k}_{\rm bond} (q_{y1}-q_{y2})^2
\partial\theta\partial\bar\theta \ ,
\end{equation}

\noindent if $q_{y1}\neq q_{y2}$---i.e. if the bond
defect has a vertical component. On the other hand,
if the bond defect is purely horizontal, it
is represented by

\begin{equation}
{2\over\pi} \tilde{k}_{\rm bond}
(q_{y1}+1)^2 (q_{x1}-q_{x2})^2 
\partial^2\theta\partial^2\bar\theta
\end{equation}

\noindent A purely horizontal bond along an open boundary 
is represented by a dimension four operator.

For closed boundaries, the bond defect is represented by

\begin{equation}
-{2\over\pi} \tilde{k}_{\rm bond} (q_{x1}-q_{x2})^2 
\partial\theta\partial\bar\theta \ ,
\end{equation}

\noindent if the bond
has a horizontal component. 
If the bond is purely vertical, 
it is represented by

\begin{equation}
-{{\tilde{k}_{\rm bond}} \over {2\pi}} 
(q_{y1}(q_{y1}+1)-q_{y2}(q_{y2}+1))^2
\partial^2\theta\partial^2\bar\theta
\end{equation}

\noindent Along closed boundaries, 
it is the
purely vertical bonds that
have dimension four. 

We have verified these field identifications of bond
defects with more general calculations, involving multiple
fields and multiple bond defects.
This required generalizing 
\maybeeq{\ref{eq:Green.bond}} for multiple
bond defects.
However, the generalization is straightforward, and not
particularly instructive, so is not shown here.

%-----------------------------------------------------------------%

\section{The ASM with Dissipation}
\label{sec:Dissipation.General}

We now consider the addition of dissipation. As explained in 
the introduction, this takes the ASM off the critical
point, as shown by both numerical simulations, and
an exact analysis~\cite{dissip.2,DissipBreaksSOC,dissip.1}.
The toppling matrix becomes

\begin{eqnarray}
\label{eq:Massive.Delta}
\Delta_{\vec{i},\vec{j}}=
\left\{
\begin{array}{cl}
4+t & {\rm if}\ \vec{i}=\vec{j} \\
-1 & {\rm if\ \vec{i}\ and\  \vec{j}\ are\ nearest\
neighbors} \\
0 & {\rm otherwise}
\end{array} 
\right.
\label{eq:Delta.dissip}
\end{eqnarray}

\noindent Now, with each toppling in the bulk of the
sandpile, $t$ grains of sand are lost. 

If we set $t=0$, we get back the original, critical
ASM. It should be noted 
that the interpretation of the continuous
$t\rightarrow 0$ limit 
is potentially problematic.
The modification to $\bf\Delta$
described in \maybeeq{\ref{eq:Delta.dissip}}
is only sensible for integer
$t$. It can be extended to rational 
$t$~\cite{DissipBreaksSOC}.
However, for $t$ rational, but not an integer,
 the interpretation of the
sandpile modifications (i.e. the $\bf B$ matrix)
associated with a WAC is changed, so that what is meant
by taking a limit of infinitesimally small, rational
$t$ is unclear. However, presumably the fact that 
the massive results are good for all integers
$t>0$ can justify an analytic continuation to
$t=0$.  Regardless, we are certainly able to 
formally expand all correlation functions
in Taylor series about $t=0$,
which is what we do here.

$t$ defines an effective mass $M$ for the sandpile,
where $t=a^2M^2$~\cite{Mahieu.Ruelle}.
$a$ can be thought of as the lattice spacing.
In looking at off-critical correlations
of LBMs, we are interested in 
correlation functions where
the number of
lattice spacings between any two of the LBMs 
is $\mathcal{O}(r/a)$. 
Taking the $a\rightarrow 0$ limit then defines
the way in which we simulataneously 
take $t\rightarrow 0$ and
distances between LBMs to 
infinity.

For the off-critical sandpile,
as discussed in~\cite{Mahieu.Ruelle},
we can use the same
methods as before to calculate correlations, with
two modifications. First, we need to use a different
$\bf B$ matrix than before. 
Previously, we required that our sandpile modifications
be conservative, which meant that the each row and each
column of $\bf B$ summed up to zero. However, for the
massive sandpile we will often want to consider
nonconservative $\bf B$'s. 
(For the unit height variable,
the sandpile
modification in \maybeeq{\ref{eq:B.Unit.Height}}
will no longer 
restrict the height of $\vec{i}$ to 1,
but rather to any height from 1 to $1+t$.)
We are most
interested in LBMs associated with WACs.
If we want to 
force the heights to the heights of the
WAC, the $\bf B$ matrix must 
be changed to

\begin{eqnarray}
{\bf B} = {\bf B}_c - t {\bf B}_{nc}
\label{eq:B.dissip}
\end{eqnarray}

\noindent ${\bf B}_c$ is the $\bf B$ matrix
that would be used for this WAC for 
the non-dissipative
ASM (e.g. \maybeeq{\ref{eq:B.Unit.Height}}), and 

\begin{eqnarray}
(B_{nc})_{\vec{i},\vec{j}}=
\left\{
\begin{array}{ll}
1 & {\rm if\ \vec{i}=\vec{j},\  and\ \vec{i}\ is\ in\ the\ WAC} \\
 &{\rm height\ configuration} \\
0 & {\rm otherwise}
\end{array}
\right.
\label{eq:B.nc}
\end{eqnarray}

\noindent By ``in the WAC height configuration,''
we mean in the set of sites whose heights
are fixed, and not in one of the bordering
sites needed to to form the $\bf B$ matrix
(e.g., for the
unit height configuration, only one site
is ``in the WAC height configuration'').
More generally, for other LBMs, ${\bf B}_{nc}$
is the nonconservative part of the $\bf B$
matrix.

Second, we need to use a new Green function 
between lattice sites. As always, the
Green function is given by the inverse of the
toppling matrix. In 
appendix~\ref{sec:Massive.Green}, 
we calculate the Green function in the limit
$t\rightarrow 0$, when the distance
between two sites scales as $1/\sqrt{t}$,
and find that it approaches $(1/2\pi)K_0(r)$,
where $K_0$ is the modified Bessel function
of the second kind. The asymptotic expansion in
$1/\sqrt{t}$ is then

\begin{eqnarray}
\nonumber
G_0 \left({{r \cos\phi}\over{\sqrt{t}}},
{{r \sin\phi}\over{\sqrt{t}}}\right) 
& \rightarrow &
{1\over{2\pi}} K_0(r) +
\sqrt{t} f_1 (r,\phi) \\
& & 
+ t f_2(r,\phi) + \mathcal{O}(t^{3/2})
\label{eq:G.K0.Expansion}
\end{eqnarray}

\noindent We have not calculated the $f$ functions, 
because they turn out to not
affect the universal parts of any correlation functions.
In principle, they can contain bounded functions of
$1/\sqrt{t}$, such as $e^{ir/\sqrt{t}}$, but we have
not explicitly indicated this $t$-dependence,
since it does not affect our analysis.
(Mahieu and Ruelle found,
for $\phi=0$ and $\phi=\pi/4$,
\maybeeq{\ref{eq:G.K0.Expansion}}, and the 
specific forms of $f_1$ and $f_2$~\cite{Mahieu.Ruelle}.
They found that $f_1=0$ for these angles,
so it is possible that
$f_1=0$ for all $\phi$, although we have not 
investigated this.)

We decompose ${\bf G}_{uv}$ as
a sum of four $N_u\times N_v$ matrices:

\begin{equation}
{\bf G}_{uv}=
{\bf G}_{uv,{\rm J}}+
{\sqrt t}{\bf G}_{uv,{\rm row}}+
{\sqrt t}{\bf G}_{uv,{\rm col}}+
t{\bf G}_{uv,{\rm both}}
\label{eq:G.dissip.general}
\end{equation}

\noindent ${\bf G}_{uv,{\rm J}}$ is a matrix in which
every element is identical.
${\bf G}_{uv,{\rm row}}$ and ${\bf G}_{uv,{\rm col}}$
are matrices in which the elements depend only
on the row index, or only on the column
index. Parts of ${\bf G}$ 
which cannot be written in these
forms go into ${\bf G}_{uv,{\rm both}}$.
All four of these matrices are  Taylor
series in $\sqrt{t}$, whose $\mathcal{O}(1)$
terms depends only on the Taylor expansion
of $(1/2\pi)K_0(r)$, and whose higher 
order terms in $\sqrt{t}$ depend on
$f_1$, $f_2$, etc. . .
For example every element of ${\bf G}_{uv,{\rm J}}$
is $(1/2\pi) K_0 (r_{uv})+\sqrt{t}
f_1(r_{uv},\phi_{uv})+\dots$.
The elements of ${\bf G}_{uv,{\rm row}}$ and
${\bf G}_{uv,{\rm col}}$ depend on only
the coordinates in the $u^{\rm th}$ LBM,
or on only the coordinates in the $v^{\rm th}$
LBM, and thus require one derivative
(finite difference)
of the Green function. The elements of
${\bf G}_{uv,{\rm both}}$ require two or more
derivatives of the Green function.

For correlations at the critical point, we saw
that ${\bf G}_{uv,{\rm J}}$, ${\bf G}_{uv,{\rm row}}$, 
and ${\bf G}_{uv,{\rm col}}$ could all be ignored
in calculating LBM correlations. However,
the arguments there relied on the fact
that every row and every column of every ${\bf B}_u$
summed to zero. That no longer holds
here, and we thus need to reconsider
which terms of the Green function we need to
keep.

We expand the 
correlation functions in powers of $t$, and
look for the lowest, nonzero, power of $t$.
As with critical correlations, 
to prove the validity of
\maybeeq{\ref{eq:General.Trace.Formula}},
we need to show that
the lowest-order
term of the $n$-point function only comes from
the parts of $\det (\id+{\bf BG})$
with $n$ terms off the block diagonal.
However, the proof is much harder
in this case; we sketch the proof
in appendix~\ref{sec:Massive.Trace.Formula}.

Similarly to \maybeeq{\ref{eq:Green.r2.expansion}},
we need to Taylor expand

\begin{equation}
G_0 \left(
{1\over{\sqrt{t}}} r_{uv}\cos\phi_{uv} + k_2-k_1,
{1\over{\sqrt{t}}} r_{uv}\sin\phi_{uv} + l_2-l_1 
\right)
\end{equation}

\noindent Then, just as 
in \maybeeq{\ref{eq:G.decompose}},
we can write ${\bf G}_{uv}$
as a sum of terms, each of which is the product of a 
length $N_u$ column vector and a length $N_v$
row vector. Defining $\vec{1}_u$ to be the
vector of length $N_u$, all of whose entries are
1, we have

\begin{widetext}

\begin{eqnarray}
\nonumber
2\pi {\bf G}_{uv} & = & K_0(r_{uv}) \vec{1}_u \vec{1}_v^T \\
\nonumber
& & + 
\left(K_0 '(r_{uv}) \cos\phi_{uv} \sqrt{t}\right)
\left(\vec{1}_u\vec{k}_v^T-\vec{k}_u\vec{1}_v^T\right)
+ \left(K_0 '(r_{uv}) \sin\phi_{uv} \sqrt{t}\right)
\left(\vec{1}_u\vec{l}_v^T-\vec{l}_u\vec{1}_v^T\right) \\
\nonumber
& & -
\left(K_0(r_{uv}) t\right)
\left( \sin^2\phi_{uv}\vec{k}_u\vec{k}_v^T 
 +\cos^2\phi_{uv}\vec{l}_u\vec{l}_v^T
 +{1\over 2}
\sin(2\phi_{uv})\left(\vec{k}_u\vec{l}_v^T+\vec{l}_u\vec{k}_v^T
\right)\right) 
\\
& & +
\left(K_0 ''(r_{uv})t\right)
\left(
\cos(2\phi_{uv})
\left(\vec{l}_u\vec{l}_v^T-\vec{k}_u\vec{k}_v^T\right)-
\sin(2\phi_{uv})
\left(\vec{k}_u\vec{l}_v^T+\vec{k}_v\vec{l}_u^T\right)
\right)
+\dots
\end{eqnarray}

\noindent In the ellipses, we have dropped not only terms
of $\mathcal{O}(t^{3/2})$ and higher, but all terms 
with $f_1$ or $f_2$. The terms with $f_1$
or $f_2$ are not necessarily higher-order in $t$ than the
terms shown. $f_1$ contributes
terms of $\mathcal{O}(\sqrt{t})$
to $\vec{1}_u\vec{1}_v^T$,
and terms of $\mathcal{O}(t)$ to
$\vec{1}_u\vec{k}_v^T$, $\vec{1}_u\vec{l}_v^T$,
$\vec{k}_u\vec{1}_v^T$, and $\vec{l}_u\vec{1}_v^T$.
$f_2$ contributes terms of $\mathcal{O}(t)$
to $\vec{1}_u\vec{1}_v^T$.
Similarly to \maybeeq{\ref{eq:Crit.N.matrix}},
we represent ${\bf G}_{uv}$
with a $3\times 3$ matrix, ${\bf N}_{uv}'$, where each element of
${\bf N}_{uv}'$ 
represents a different choice of row vector and
column vector:

\pagebreak

\begin{eqnarray}
\nonumber
& & \qquad\qquad\qquad\qquad\qquad
\vec{k}_v^T \qquad\qquad\qquad\qquad\qquad\qquad\qquad
\vec{l}_v^T \qquad\qquad\qquad\qquad\qquad\qquad\qquad
\vec{1}_v^T
\\
{\bf N}_{uv}' & \equiv &
{1\over{2\pi}}
\left(
\begin{array}{ccc}
-t(\sin^2\phi K_0+\cos(2\phi)K_0'')+\mathcal{O}(t^{3/2}) &
(t/2)\sin(2\phi)(K_0-2K_0''))+\mathcal{O}(t^{3/2}) &
-K_0' \cos\phi \sqrt{t}+\mathcal) \\
(t/2)\sin(2\phi)(K_0-2K_0'')+\mathcal{O}(t^{3/2}) &
-t(\cos^2\phi K_0-\cos(2\phi)K_0'')+\mathcal{O}(t^{3/2}) &
-K_0' \sin\phi \sqrt{t}+\mathcal{O}(t) \\
K_0' \cos\phi \sqrt{t}+\mathcal{O}(t) &
K_0' \sin\phi \sqrt{t}+\mathcal{O}(t) &
K_0+\mathcal{O}(t^{1/2})
\end{array} 
\right)  
\begin{array}{c}
\vec{k}_u \\
\vec{l}_u \\
\vec{1}_u
\end{array}
\nonumber\\
\label{eq:N.dissip}
\end{eqnarray}

\end{widetext}

\noindent To save space, we have here abbreviated
$\phi_{uv}\rightarrow\phi$ and
$K_0(r_{uv})\rightarrow K_0$.
Now, while 
some of the terms depending on $f_1$ and $f_2$
are $\mathcal{O}(\sqrt{t})$ and $\mathcal{O}(t)$,
they are higher order terms
in ${\bf G}_{uv,{\rm J}}$, ${\bf G}_{uv,{\rm row}}$,
${\bf G}_{uv,{\rm col}}$, and ${\bf G}_{uv,{\rm both}}$,
and thus are higher order 
in the specific matrix elements
of ${\bf N}_{uv}'$ that they contribute to.
We will later see that this justifies dropping
them.

Just as in section~\ref{sec:WAC.npoint}, when each
off-diagonal Green function matrix is replaced by the product of a
column vector and a row vector, each
$(\id+{\bf B}_u{\bf G}_{uu})^{-1}{\bf B}_u$
is bracketed by a row vector to its left and a column 
vector to its right, producing a $1\times 1$ matrix.
We thus get a matrix, ${\bf M}_u'$,
similar to the ${\bf M}_u$ of
\maybeeq{\ref{eq:Crit.M.matrix}}.  
${\bf M}_u'$ is $3\times 3$, 
rather than $2\times 2$, because
the vectors bracketing 
$(\id+{\bf B}_u{\bf G}_{uu})^{-1}{\bf B}_u$
to the left and to the right can now be
$\vec{k}_u$, $\vec{l}_u$, or $\vec{1}_u$.

In principle, when calculating ${\bf M}_u'$,
${\bf B}_u$ and ${\bf G}_{uu}$ 
should be the matrices for the massive sandpile.
However, we only need most elements of
${\bf M}_u'$ in the limit $t\rightarrow 0$ (this will
be justified shortly).
In this limit, we
replace ${\bf B}_u$ 
with ${\bf B}_{u,c}$, and the 
elements of ${\bf G}_{uu}$ with the normal, well-known,
critical Green function. 
Thus, to lowest-order, 
the elements of ${\bf M}_u$ in ${\bf M}_u'$ are unchanged:

\begin{eqnarray}
c_{u,kk}' & = & c_{u,kk} + \mathcal{O}(t) \\
c_{u,kl}' & = & c_{u,kl} + \mathcal{O}(t) \\
c_{u,ll}' & = & c_{u,ll} + \mathcal{O}(t) 
\end{eqnarray}

\noindent For the new entries of ${\bf M}_u'$,
it is not hard to show that

\begin{eqnarray}
c_{u,1k}' & \equiv &
-p_u
\vec{1}_u^T {\id\over{\id+{\bf B}_u{\bf G}_{uu}}}
{\bf B}_u \vec{k}_u = \mathcal{O}(t) \\
c_{u,1l}' & \equiv &
-p_u
\vec{1}_u^T {\id\over{\id+{\bf B}_u{\bf G}_{uu}}}
{\bf B}_u \vec{l}_u = \mathcal{O}(t) \\
\nonumber
c_{u,11}' & \equiv &
-p_u
\vec{1}_u^T {\id\over{\id+{\bf B}_u{\bf G}_{uu}}}
{\bf B}_u \vec{1}_u \\
\nonumber
& = & -p_u
\vec{1}_u^T (-t {\bf B}_{u,nc}) \vec{1}_u  +
\mathcal{O}(t^2) \\
& = & tp_u C +\mathcal{O}(t^2) \ ,
\end{eqnarray}

\noindent where $p_u$ is the probability for the 
LBM, and C is defined as
the number of sites in the WAC
height configuration (as defined
below \maybeeq{\ref{eq:B.nc}}).
$c_{u,11}'$ is the only element of ${\bf M}_u'$
where we need the $\mathcal{O}(t)$ term.
Mahieu and Ruelle defined $C$ as the coefficient 
in \maybeeq{\ref{eq:LCFT.rep}},
and observed that for the 14
WACs that they considered, $C$ always turned out to
always be equal to the number of sites in the 
WAC height configuration~\cite{Mahieu.Ruelle}.
We will see that our $C$ is the same as their $C$,
proving that their
observation holds for
all WACs. 

We now have

\begin{eqnarray}
{\bf M}_{u}' \equiv
\left(
\begin{array}{ccc}
c_{u,kk}+\mathcal{O}(t) &
c_{u,kl}+\mathcal{O}(t) &
\mathcal{O}(t) \\
c_{u,kl}+\mathcal{O}(t) &
c_{u,ll}+\mathcal{O}(t) &
\mathcal{O}(t) \\
\mathcal{O}(t) &
\mathcal{O}(t) &
tp_u C +\mathcal{O}(t^2)
\end{array} 
\right)  
\nonumber\\
\label{eq:M.dissip}
\end{eqnarray}

\noindent The $n$-point correlation is given by

\begin{eqnarray}
\nonumber
-\trace\left( {\bf M}_{u_1}' {\bf N}_{u_1u_2}'
{\bf M}_{u_2}' {\bf N}_{u_2u_3}' \dots
{\bf M}_{u_n}' {\bf N}_{u_nu_1}'\right) & & \\
& & \hspace{-2.3in}
- \left( ((n-1)!-1)\ {\rm other\ trace\ terms} \right)
\label{eq:Massive.npt.Trace.Form}
\end{eqnarray}

\noindent It is easy to now verify that the
higher-order
terms in $t$ in ${\bf M}_u'$ and ${\bf N}_{uv}'$
that we dropped in \maybeeq{\ref{eq:N.dissip}}
and \maybeeq{\ref{eq:M.dissip}}
indeed give contributions
of $\mathcal{O}(t^{n+1/2})$
or higher to the $n$-point correlation, justifying
the approximations used. We
now have all correlations of LBMs in the ASM
with dissipation.

We can show that the correlations we have found are
the same as the correlations of the field in
\maybeeq{\ref{eq:LCFT.rep}}. Mahieu and Ruelle 
proposed that the appropriate massive
extension of the LCFT in
\maybeeq{\ref{eq:LCFT.action}} is~\cite{Mahieu.Ruelle}

\begin{equation}
S= {1\over\pi} \int d^2x
\left\{ :\partial\theta\bar\partial\bar\theta:+
{M^2\over 4} : \theta\bar\theta :
\right\}
\end{equation}

\noindent This theory is still Gaussian, with correlation
functions

\begin{eqnarray}
\left< \theta(z_u)\bar\theta(z_v) \right> & = &
K_0 \left( M  |z_u-z_v|  \right) \\
\left< \theta(z_u)\theta(z_v) \right> & = & 0 \\
\left< \bar\theta(z_u)\bar\theta(z_v) \right> & = & 0
\end{eqnarray}

\noindent Other two-point correlations can then be obtained
by taking holomorphic or antiholomorphic derivatives.
Then, just as in section~\ref{sec:WAC.npoint},
since the theory is Gaussian, we can write the
$n$-point correlation of \maybeeq{\ref{eq:LCFT.rep}}
exactly. 
The result is formally identical to
\maybeeq{\ref{eq:LCFT.critical.trace}}, where the
analogues of $\bf F$ and $\bf H$
in \maybeeqs{\ref{eq:F.critical}-\ref{eq:H.critical}}
are now $3\times 3$ (since the $\theta$ and $\bar\theta$
fields in \maybeeq{\ref{eq:LCFT.rep}} may
have holomorphic derivatives, antiholomorphic
derivatives, or no derivatives at all).
Using the identifications in 
\maybeeq{\ref{eq:A}-\ref{eq:B2}}, and $t=a^2M^2$,
the results agree with the correlation function
in \maybeeq{\ref{eq:Massive.npt.Trace.Form}}
(up to a proportionality factor, $a^{2n}$).

%-----------------------------------------------------------------%

\acknowledgments{We thank P. Ruelle for useful discussions.}

%-----------------------------------------------------------------%

\appendix

\section{Massive Green Function}
\label{sec:Massive.Green}

The Green function for the ASM with dissipation is
the inverse of the toppling matrix $\bf\Delta$, given
in \maybeeq{\ref{eq:Massive.Delta}}. In the limit of
an infinite lattice, this can be found
by Fourier transform~\cite{Mahieu.Ruelle}:

\begin{equation}
G_0(m,n) = 
\int_{-\pi}^{\pi} {{dp_x}\over{2\pi}}
\int_{-\pi}^{\pi} {{dp_y}\over{2\pi}}
{{e^{i(p_xm+p_yn)}}\over{4+t-2\cos p_x-2\cos p_y}}
\end{equation}

\noindent We are interested in this integral either
for $m$ and $n$ both $\mathcal{O}(1)$,
or for $m$ and $n$ both proportional to $1/\sqrt{t}$,
in the limit $t\rightarrow 0$.

We saw in section~\ref{sec:Dissipation.General},
that for $m$ and $n$ both $\mathcal{O}(1)$,
we only needed $G(m,n)-G(0,0)$,
in the limit $t\rightarrow 0$. But this is just the
standard, dissipation-free, lattice Green function,
whose properties can be looked up in standard
references---for example,~\cite{Spitzer}.
So we only need to discuss the case where $m$ and $n$
are both proportional to $1/\sqrt{t}$:

\begin{eqnarray}
\nonumber
\lim_{t\rightarrow 0} 
G_0 \left( 
{{r\cos\phi}\over{\sqrt{t}}},
{{r\sin\phi}\over{\sqrt{t}}}
\right) 
& = & \\
& & \hspace{-1.6in}
\lim_{t\rightarrow 0} 
\int_{-\pi}^{\pi} {{dp_x}\over{2\pi}}
\int_{-\pi}^{\pi} {{dp_y}\over{2\pi}}
{{e^{i(p_x\cos\phi+p_y\sin\phi)r/\sqrt{t}}}\over
 {4+t-2\cos p_x-2\cos p_y}}
\end{eqnarray}

\noindent In the limit $t\rightarrow 0$, 
the exponential oscillates infinitely rapidly, so when
multiplied by any function smooth in the limit
$t\rightarrow 0$, gives an integral of zero. So no
error is introduced by changing the region of
integration to a small disc of radius $\epsilon$
about $(p_x,p_y)=(0,0)$. Similarly, because

\begin{equation}
{1\over{4+t-2\cos p_x-2\cos p_y}}-
{1\over{p_x^2+p_y^2+t}}
\end{equation}

\noindent is smooth over this region, we can replace the
first term with the second one.
Changing to polar coordinates,
the integral is

\begin{equation}
\lim_{t\rightarrow 0} 
\int_0^\epsilon p dp 
\int_0^{2\pi} d\alpha
{{e^{i(\cos\phi\cos\alpha+\sin\phi\sin\alpha)rp/\sqrt{t}}}\over
{(2\pi)^2(p^2+t)}}
\end{equation}

\noindent With a change of variables, the integral becomes

\begin{eqnarray}
\nonumber
\lim_{t\rightarrow 0} 
G_0 \left( 
{{r\cos\phi}\over{\sqrt{t}}},
{{r\sin\phi}\over{\sqrt{t}}}
\right) 
& = & \\
\nonumber
& & \hspace{-1.0in} =
{1\over{(2\pi)^2}}
\int_0^\infty {{p dp}\over{p^2+1}}
\int_0^{2\pi} d\alpha e^{irp\cos\alpha} \\
& & \hspace{-1.0in} =
{1\over{2\pi}} K_0(r) \ ,
\end{eqnarray}

\noindent where $K_0$ is the modified Bessel function of the
second kind.

To find the leading-order critical limit, we take
$r\rightarrow 0$ (while $r/\sqrt{t}$ is still large), and
use $K_o(r)\rightarrow -\ln (r)$, reproducing the first
term of \maybeeq{\ref{eq:Bulk.Green}}.

%-----------------------------------------------------------------%

\section{The trace formula for massive correlations}
\label{sec:Massive.Trace.Formula}

In this appendix, we sketch the proof that the trace formula
in \maybeeq{\ref{eq:General.Trace.Formula}}
is valid for all $n$-point correlations of LBMs
off the critical point.
This requires showing that in the determinant
$\det (\id+{\bf BG})$, if we only want terms up
to $\mathcal{O}(t^n)$, we never need to pick
more than $n$ terms off the block diagonal of
$\id+{\bf BG}$. 
In other words, when we calculate the 
determinant with

\begin{equation}
\det{\bf X} = 
\sum_{p\in S_{\mid X\mid}} (-)^p X_{1,p(1)} X_{2,p(2)} \dots
X_{n,p(n)}\ ,
\label{eq:determinant.definition}
\end{equation}

\noindent where $p$ is summed over all permutations of 
$\{1,2,\dots ,|X| \}$,
we never have more than $n$ of the $X_{i,p(i)}$'s
from the off-diagonal blocks.
The off-diagonal blocks of $\id+{\bf BG}$ all
have the form ${\bf B}_u{\bf G}_{uv}$, $u\neq v$,
where ${\bf B}_u$ and ${\bf G}_{uv}$ can be written
with \maybeeq{\ref{eq:B.dissip}} 
and \maybeeq{\ref{eq:G.dissip.general}}. Since
every row of ${\bf B}_{c,u}$ sums to zero, 
${\bf B}_{c,u}{\bf G}_{uv,J}={\bf 0}$, and
${\bf B}_{c,u}{\bf G}_{uv,{\rm col}}={\bf 0}$.
The off-diagonal block can thus be written

\begin{eqnarray}
\nonumber
{\bf B}_u{\bf G}_{uv} & = &
\sqrt{t} {\bf B}_{c,u} {\bf G}_{uv,{\rm row}} +
t {\bf B}_{c,u} {\bf G}_{uv,{\rm both}} \\
& & 
- t {\bf B}_{nc,u} {\bf G}_{uv,J}
+ \mathcal{O}(t^{3/2})
\label{eq:massive.G.offdiag.expansion}
\end{eqnarray}

\noindent (We need the $\mathcal{O}(t^{3/2})$
terms for the calculation of the correlation
functions in section~\ref{sec:Dissipation.General},
but do not need their explicit form for this proof.)
Since there are terms of $\mathcal{O}(\sqrt{t})$
in the off-diagonal blocks, naively, to get the 
$\mathcal{O}(t^n)$ contribution to the correlation function,
we would need parts of the determinant with up to 
$2n$ terms off the block diagonal. So we need to explain
why the terms with more than $n$ terms off the block
diagonal in fact have all their contributions to
the $\mathcal{O}(t^n)$ part of the correlation
function cancel (as well as why all the terms with
lower powers of $t$ cancel).

We define a ``row matrix'' to be a matrix in which 
the entries depend only on the row index 
When we consider
contributions to $\det (\id+{\bf BG})$, 
we consider,
for each contributing matrix element off the 
block diagonal, whether it is from
${\bf B}_{c,u}{\bf G}_{uv,{\rm row}}$,
${\bf B}_{c,u}{\bf G}_{uv,{\rm both}}$,
${\bf B}_{nc,u}{\bf G}_{uv,{\rm J}}$,
or from the elements of $\mathcal{O}(t^{3/2})$
or higher.
We use several matrix theorems; the first two are 
general, not referring specifically to $\bf B$
or $\bf G$.
It is not hard to show

\begin{theorem}
The determinant has zero contribution from terms
that have both a matrix element from a row matrix in the 
$(u,v_1)$ block, and a matrix element from a row
matrix in the $(u,v_2)$ block.
\label{thm:1}
\end{theorem}

\noindent By ``zero contribution,'' we mean that while
specific terms in \maybeeq{\ref{eq:determinant.definition}}
may be nonzero, when all such terms
are considered, they cancel. 
Theorem~\ref{thm:1} is easy to prove: if
$X_{\alpha\beta}$ is the element from the $(u,v_1)$ block,
and $X_{\gamma\delta}$ is the element from the
$(u,v_2)$ block, we get a cancelling contribution 
from $X_{\alpha\delta}$ and $X_{\gamma\beta}$.
It is somewhat harder to show the following theorem,
which we state without proof:

\begin{theorem}
Suppose the $(u,v)$ block ($u\neq v$) is a row matrix, 
every column of the $(v,v)$ block sums to one, and
for every $u'\neq v$,
every column of the $(v,u')$ block
sums to zero. Then the determinant has
zero contribution from the matrix elements of the $(u,v)$
block; in other words, the determinant is unchanged
if every element of the $(u,v)$ block is set to zero.
\label{thm:2}
\end{theorem}

We note that
${\bf B}_{c,u}{\bf G}_{uv,{\rm row}}$ and
${\bf B}_{nc,u}{\bf G}_{uv,{\rm J}}$
are both row matrices, and 
that every column of ${\bf B}_{c,u}{\bf G}_{uv,{\rm row}}$
sums to zero, as does every column of
${\bf B}_{c,u}{\bf G}_{uv,{\rm both}}$.
Then, repeatedly
applying theorems~\ref{thm:1} and~\ref{thm:2}
allows us to prove the following:

\begin{theorem}
In the determinant $\det (\id+{\bf BG})$, 
suppose the ($u_0,u_1$) block ($u_0\neq u_1$)
has a contributing matrix element
from ${\bf B}_{c,u_0}{\bf G}_{u_0u_1,{\rm row}}$.
Then, in the terms that produce a nonzero contribution
to the determinant, there is an ordered sequence of
distinct block indices, $(u_1,u_2,\dots,u_x)$,
$x\geq 1$, such that for all $1\leq i< x$, the
$(u_i,u_{i+1})$ block has a term from 
$-t {\bf B}_{nc,u_i}{\bf G}_{u_i u_{i+1},J}$.
Furthermore, either \\
1) The $(u_x,u_x)$ block has a term of $\mathcal{O}(t)$
or higher. \\
or \\
2) There is a term of order $\mathcal{O}(t^{3/2})$
or higher in an off-diagonal block, $(u_x,u_{x+1})$. \\
A different $(v_0,v_1)$ block
$(v_0\neq v_1)$
with a matrix element from
${\bf B}_{c,v_0}{\bf G}_{v_0v_1,{\rm row}}$
will produce an ordered sequence
of distinct block indices,
$(v_1,v_2,\dots v_y)$, with no elements
in common with $(u_1,u_2,\dots u_x)$.
\end{theorem}

Next, for any nonvanishing contribution to the determinant, we
define

\begin{eqnarray}
c_1 & = & \#\ {\rm terms\ off\ the\ block\ diagonal} \\
\nonumber
c_2 & = & \#\ {\rm terms\ off\ the\ block\ diagonal\ that}\\
& & {\rm are\ exactly\ \mathcal{O}(t^{1/2})} \\
\nonumber
c_3 & = & \#\ {\rm terms\ off\ the\ block\ diagonal\ that}\\
& & {\rm are\ \mathcal{O}(t^{3/2})\ or\ higher} \\
\nonumber
c_4 & = & \#\ {\rm terms\ on\ the\ block\ diagonal\ that}\\
& & {\rm are\ \mathcal{O}(t)\ or\ higher}
\end{eqnarray}

\noindent Theorem 3 shows that each term in the
determinant of type $c_2$ can be associated with a distinct
term of type $c_3$ or $c_4$, so that $c_3+c_4\geq c_2$.
The number of powers of $t$ from 
this contribution to the
determinant is then {\it at least}

\begin{equation}
(c_1-c_2-c_3)(1)+c_2(1/2)+c_3(3/2)+c_4(1)\geq c_1
\label{eq:order.inequality}
\end{equation}

\noindent So if we only want $\mathcal{O}(t^n)$
contributions to the correlation function we should never
have more than $n$ terms off the block diagonal. Furthermore,
we want a connected correlation function, so we should
always have exactly $n$ terms off the block diagonal.
This concludes the proof that
\maybeeq{\ref{eq:General.Trace.Formula}}
is valid for off-critical $n$-point correlations.

To get strict equality in
\maybeeq{\ref{eq:order.inequality}},
we need $c_1=n$, $c_4=0$, and $c_2=c_3$.
Furthermore, the terms 
of type $c_3$ should be
exactly proportional to $t^{3/2}$.
The fact that $c_4=0$ means that in the diagonal blocks,
$\id+{\bf B}_u{\bf G}_{uu}$,
we can set $t=0$ at the start of our calculations,
as already seen by other means in
section~\ref{sec:Dissipation.General}.

%-----------------------------------------------------------------%

\end{document}